\documentclass{article}

\usepackage{PRIMEarxiv}

\usepackage[utf8]{inputenc} 
\usepackage[T1]{fontenc}    
\usepackage{hyperref}       
\usepackage{url}            
\usepackage{booktabs}       
\usepackage{amsfonts}       
\usepackage{nicefrac}       
\usepackage{microtype}      
\usepackage{lipsum}
\usepackage{fancyhdr}       
\usepackage{graphicx}       
\usepackage{amsmath}        
\usepackage{amssymb}        
\usepackage{siunitx}        
\usepackage{float}          
\usepackage{caption}        

\title{Optimizing Quantum Key Distribution Network Performance using Graph Neural Networks
}

\author{
  Akshit Pramod Anchan, Ameiy Acharya, and Leki Chom Thungon \\
  School of Computer Science and Engineering (SCOPE) \\
  Vellore Institute of Technology \\
  Chennai - 600127, India \\
  \texttt{lekichom.thungon@vit.ac.in} \\
}

\begin{document}
\maketitle

\begin{abstract}
This paper proposes an optimization of Quantum Key Distribution (QKD) Networks using Graph Neural Networks (GNN) framework. Today, the development of quantum computers threatens the security systems of classical cryptography. Moreover, as QKD networks are designed for protecting secret communication, they suffer from multiple operational difficulties: adaptive to dynamic conditions, optimization for multiple parameters and effective resource utilization. In order to overcome these obstacles, we propose a GNN-based framework which can model QKD networks as dynamic graphs and extracts exploitable characteristics from these networks' structure. The graph contains not only topological information but also specific characteristics associated with quantum communication (the number of edges between nodes, etc). Experimental results demonstrate that the GNN-optimized QKD network achieves a substantial increase in total key rate (from 27.1 Kbits/s to 470 Kbits/s), a reduced average QBER (from 6.6\% to 6.0\%), and maintains path integrity with a slight reduction in average transmission distance (from 7.13 km to 6.42 km). Furthermore, we analyze network performance across varying scales (10 to 250 nodes), showing improved link prediction accuracy and enhanced key generation rate in medium-sized networks. This work introduces a novel operation mode for QKD networks, shifting the paradigm of network optimization through adaptive and scalable quantum communication systems that enhance security and performance.
\end{abstract}


\section{Introduction}
\label{sec:introduction}
The rapid advancement of quantum computing presents a significant and unprecedented threat to the security of contemporary cryptographic systems. Many widely used encryption methods rely on computational complexity, a foundation that quantum computers are poised to dismantle \cite{b601}. This vulnerability requires transferring to cryptographic techniques providing information-theoretic security that ensures secure information even with unlimited computational capabilities of the adversary \cite{b1}. The QKD is a prospective technique presenting a solution for the described issue. The QKD uses basic laws of quantum mechanics for establishing secure keys between the communicating parties ensuring that any eavesdropping can be detected as any such action will lead to unavoidable disturbances \cite{b2}.

The fundamental theory of QKD is based on the no-cloning theory and Heisenberg's uncertainty theory, both of which are roots of quantum mechanics. These theories allowed QKD to achieve a secure key exchange between both parties \cite{b3}. The no-cloning theorem is one of the essential theorems of quantum information theory and states that it is impossible to create an identical copy of an arbitrary unknown quantum state without disturbing the original state. This unique property determines that any eavesdropper's attempts to intercept the key will result in the introduction of disturbances in the quantum channel, which can be detected by legitimate users \cite{b602}. These disturbances will result in increased error rates in the quantum states, which can be assessed during the protocol's reconciliation process. Unlike classical cryptographic systems based on the principle of computational hardness, the key distribution's security is based on an unbreakable physical principle \cite{b4}. As a result, QKD provides information-theoretic security against brute-force attacks, which means that the security of the QKD does not depend upon the eavesdropper's computational capabilities or the progress achieved in the development of breaking algorithms.

Heisenberg's uncertainty principle is also a significant in QKD's security model. It gives boundaries to the accuracy of measuring quantum systems' complementary features at a particular moment \cite{b5}. In practical systems, qubits are encoded in non-orthogonal quantum states, putting potential eavesdroppers in an impossible predicament: they have to measure the transmitted qubits to attain information, but cannot do so without irretrievably disturbing those states \cite{b603}.

The most widely studied QKD protocol, BB84, introduced by Bennett and Brassard in 1984, utilizes quantum states—typically polarized photons—to encode information. When a sender (Alice) transmits quantum-encoded bits to a receiver (Bob), any attempt at eavesdropping by a third party (Eve) introduces detectable disturbances due to the fundamental nature of quantum measurement. This guarantees that any interception attempt is not only identified but also allows legitimate users to discard compromised key bits, ensuring the remaining shared key is secure \cite{b6}. Other protocols, such as E91 based on entanglement, further strengthen security by leveraging quantum correlations that cannot be reproduced by classical means.

\begin{figure}[h]
\centering
\fbox{\includegraphics[width=0.7\textwidth]{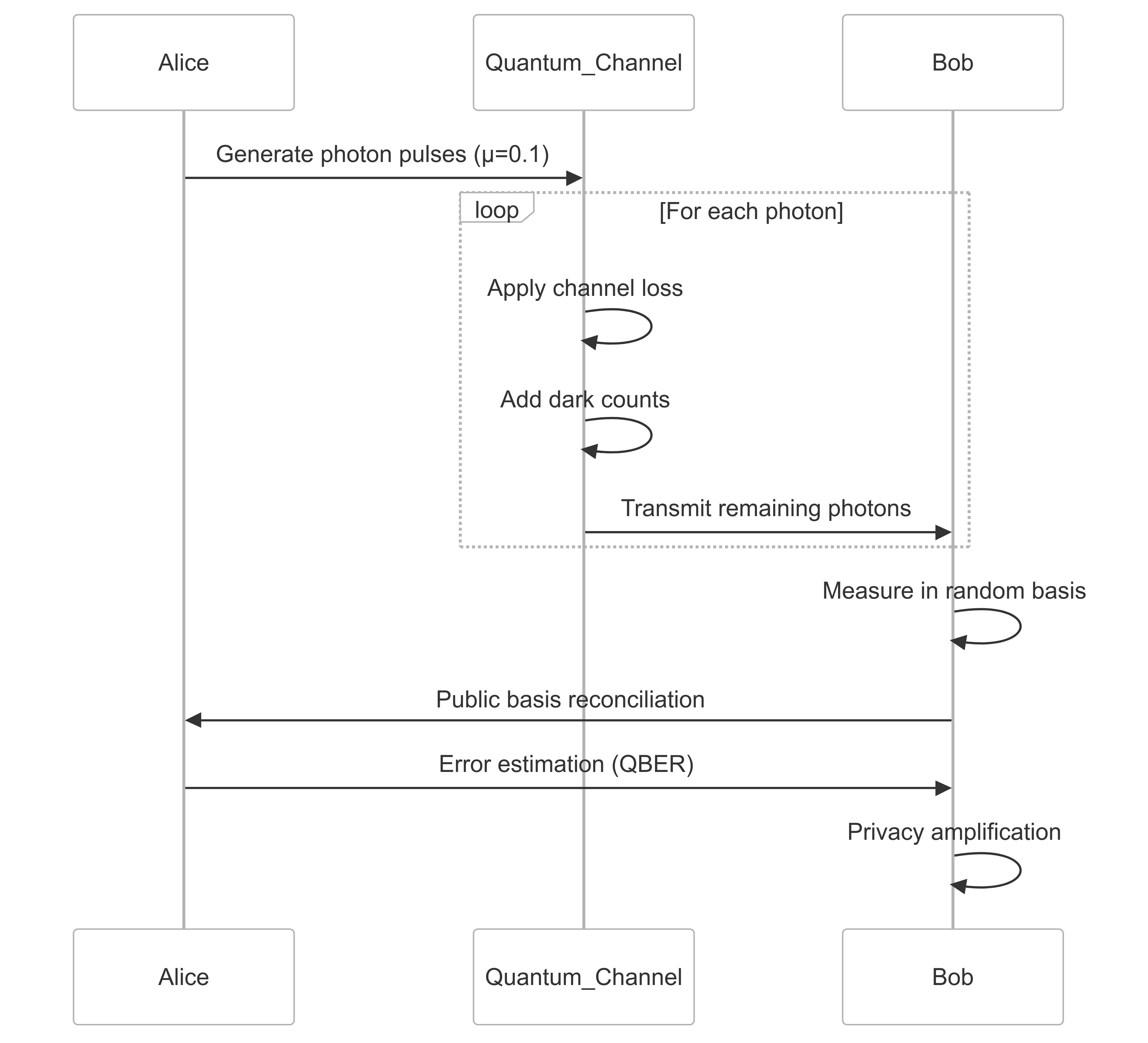}}
\caption{Simplified Demonstration of BB84 Communication Protocol}
\label{fig:arch_bb84}
\end{figure}

The BB84 quantum key distribution protocol relies on the transmission of quantum states encoded in two conjugate bases, typically the rectilinear ($+$) and diagonal ($\times$) bases for polarization-encoded photons \cite{b930}. Alice prepares each photon in one of the four states $\{|0\rangle, |1\rangle, |+\rangle, |-\rangle\}$ chosen uniformly at random, where $|+\rangle = \frac{1}{\sqrt{2}}(|0\rangle + |1\rangle)$ and $|-\rangle = \frac{1}{\sqrt{2}}(|0\rangle - |1\rangle)$. Bob measures each received photon in a randomly chosen basis ($+$ or $\times$). After the quantum transmission, they perform basis reconciliation by disclosing their basis choices over a classical channel, keeping only the bits where their bases matched. The quantum bit error rate (QBER) is estimated by comparing a subset of the remaining bits:

\begin{equation}
\text{QBER} = \frac{\text{Number of discordant bits}}{\text{Total bits compared}}
\end{equation}

Privacy amplification is then applied to distill a shorter but more secure key. The security of BB84 stems from the no-cloning theorem and the disturbance of quantum states upon measurement, which reveals eavesdropping attempts. Despite its theoretical security guarantees, practical QKD implementations face several real-world limitations \cite{b890}. The transmission of quantum states is highly susceptible to noise, loss, and decoherence, particularly over long distances in optical fiber or free-space channels. Additionally, the necessity of single-photon sources, highly sensitive detectors, and quantum-compatible infrastructure poses technical and economic challenges.

To extend the reach of QKD beyond direct point-to-point connections, solutions such as trusted-node relays and quantum repeaters have been explored \cite{b605}. However, these methods introduce trade-offs in terms of security and efficiency \cite{b719}. As a result, optimizing QKD networks for scalable and resilient operation requires sophisticated strategies that account for both classical network constraints and quantum-specific challenges.

While QKD offers robust security for point-to-point communication, its practical deployment in larger, more complex networks faces significant operational challenges. As QKD networks expand beyond simple connections to encompass metropolitan and long-distance scenarios, factors such as dynamic network adaptation to fluctuating conditions, optimization of multiple interdependent parameters (including key generation rates and error rates), scalability constraints, and efficient management of limited quantum channel resources become crucial. Traditional static routing and resource allocation algorithms, commonly employed in classical networks, struggle to handle these challenges effectively, particularly in real-time decision-making scenarios that are characteristic of dynamic QKD networks.

GNNs have been shown to have great potential for learning to solve complex network optimization tasks in a variety of contexts \cite{b819}. In this context, some works by Huang et al., Hou et al. \cite{b900} and by Shen et al. \cite{b100} have used GNNs in modeling wireless communications and networks. Similar works are presented by Ferriol-Galmes et al. \cite{b110}. GNNs are a good candidate to tackle the QKD network due to their capacity to handle graph-structured data, which the QKD networks represent. GNNs can learn complex relationships in the data represented in the graph topology. Besides, some works addressing quantum networks have explored the application of GNNs in modeling them, as in the cases of Verdon et al. \cite{b819} and Ceschini et al \cite{b120}. There have also been proposals, such as that by Qian et al. \cite{b130}, for modeling of such networks using alternate continuous variable methods.

As numerous countries progress in deploying metropolitan QKD networks and developing quantum internet infrastructure, the limitations of classical network optimization methods have become increasingly evident. The contribution of the research is a general structure for QKD network optimization based on the Graph Neural Networks with the promise of improving network productivity related to security and performance is still a breakthrough theory and practice for QKD networks. QKD networks based on GNNs will build a scalable comparative optimal communication model that can adapt and secure newly given conditions \cite{b604}. The research-scale focused on developing a GNN model that can learn and estimate the correlation between QKD network components under unique quantum conditions, including decoherence, error conditions, key rate generation, etc \cite{b140}. GNNs can consider both topological network characteristics and the quantum criteria used in the network. The proposed framework is thoroughly tested with standard measures, focusing on higher key generation rate, lower latency, and enhanced security assurances under network disturbances. A quantum-aware GNN architecture designed specifically for QKD networks opens up a new avenue for research that can further improve the adaptability, efficient resource utilization, and security-conscious decisions of the large-scale quantum communication systems \cite{b150}. As the development of a quantum internet moves forward, it becomes increasingly necessary to protect delicate communications from potential quantum-related attacks.

\section{Proposed Methodology}
\label{sec:methodology}
At the heart of the proposed methodology is a simulation environment that can faithfully recreate the behavior of a QKD network and allow the optimization of graph neural networks to be used. The simulation environment comprises several modules that can help create topologies for the QKD network, simulate the properties of quantum channels, and allow for the training of a GNN-based link prediction algorithm.

Network topologies are created using a probabilistic model in the form of multivariate normal distributions. It provides clustered distributions of nodes, closely resembling the actual deployments of networks. In real-life scenarios, nodes can be clustered at locations, for example, the distances between the nodes are calculated and links are created on the basis of predefined thresholds of distances, which in the case of QKD, specify the physical parameters of the system.

The most significant facet of the generation of the network topology is its dynamic integration process with the channel simulator. A quantum channel simulator that executes the BB84 protocol is integrated with the parameters that can be modified according to physical scenarios \cite{b607}. BB84 is a common protocol in the deployment of a majority of QKD solutions. Significant parameters influencing key generation and underlying characteristics of quantum key distribution schemes have also been integrated into the simulations such as detector efficiency, dark count rates, fiber loss, etc \cite{b606}. These factors directly associate with the quantum bit error rate (QBER) which becomes a function of the distance between nodes, as signals will typically incur more losses for larger distances \cite{b608}. The QKD simulations also take into account the statistics of the emitted photons determined through a Poisson distribution which forms a conventional model for photon statistics \cite{b609}. Lastly, the derived key rates factor in the security parameter associated with the BB84 protocol which highlights the considerations taken to represent the achievable key rates from the established QKD simulations and modeling. The simulation incorporates both free space and fiber channels for quantum key generation allowing differing implementations for quantum networks.

Finally, the output network topologies with their quantum channels properties are converted into a PyTorch Geometric Data objects that can be directly handled by graph neural network (GNN) libraries \cite{b610}. The feature of the node supposed to include position, degree (no. of edges connected), and the betweenness centrality (importance of node that connect other nodes) of the node. These node features are significant for GNN to know the topological features of the network and the position information of nodes.

For the link prediction task, we use an advanced GNN model based on up-to-date graph neural network (GNN) layers. The node features are processed in the TransformerConv and GATv2Conv layers, which enables the GNN to learn the sophisticated dependencies between nodes and node features \cite{b611}\cite{b612}. The edge features, which describe the properties of the link (also obtained from the quantum channel simulation) between nodes, are processed in a multi-layer perceptron (MLP) architecture. Finally, all the processed node features and edge features are concatenated in the model, and the key establishment success probability is predicted for each pair of nodes in the networks. It is subsequently used to optimize the resource allocation and routing. The architecture of the model is equipped with dropout and layer normalization strategies to improve the generalizability and overcome overfitting \cite{b613}. The final output produced by the model is the link probability corresponding to all possible links in the network topology.

The specific parameter configurations used throughout the simulation are as follows. The default network setup is initialized with 20 nodes, representing a base network size. A maximum distance threshold of 100 units is set for establishing links between these nodes. As for the quantum channel setting, a 1550nm wavelength is set for any relevant optical transmissions. A Fiber loss coefficient of 0.2 dB/km is configured. Detector efficiency is set to 10\%. The dark count rate is configured to be $10^{-6}$. The mean photon number per pulse is set to 0.1. Lastly, 10,000 pulses per measurement are considered. As for the training of the Machine Learning Model, 64 hidden channels are present in the neural network. A regularization dropout of 20\% is incorporated. The model is trained over 200 epochs. A 5-fold cross validation scheme is used. An early stopping patience of 20 epochs is used. To simulate the environment, the center covariance is 100 for node clustering. The node covariance is set to 10 for the local distribution. There is a 20\% probability assigned for atmospheric visibility effects. Finally, a 20km atmospheric visibility parameter is applied, where applicable.

\section{Experimental Setup}
\label{sec:experiement}
The experimental setup is designed to rigorously evaluate the performance of the proposed GNN-based QKD network optimization framework. A comprehensive suite of tools and procedures is employed to train the GNN model, assess its predictive capabilities, and analyze the resulting network characteristics. The core of the evaluation process revolves around a robust training and validation regimen.
\href{https://github.com/AmeiyAcharya/QKD_Optimisation_GNN}{The entire code is available on \underline{GitHub}.}

The training and testing of the proposed GNN model is implemented through the script \verb|main.py|. K-fold cross validation is implemented in this experiment. K-fold cross validation is where the given dataset is separated into K-subset or folds and training is applied on (K-1) folds and validation on the excluded fold \cite{b614}. This is repeated K-time and in each repetition, a fold is used as validation set. It helps to reduce overfitting by data unreliability when split and the model prediction accuracy on unseen data will shortly be better approximated as well. Early stopping is applied to the model as well to avoid overfitting \cite{b615}. The validation loss is monitored during training and if no decrease is observed after a specific number of epochs, training is stopped. It makes the model trained for an optimum time, saving unnecessary time and preventing from overfitting. The model is trained using AdamW, an advanced version of Adam optimizer that includes weight decay regularization to improve performance of deep neural networks \cite{b616}. The learning rate parameters are further adjusted using a learning rate schedule during training to reduce the convergence time.

Link prediction task is trained using negative sampling \cite{b617}. Since, the number of potential links in a network is significantly greater than the actual links, using all links in the training data is computationally inefficient and causes a bias where links are predicted to not be present. Negative sampling mitigates this by including random samples of non-existent links in the training data in addition to the existing links. The training data becomes more balanced and helps the model to predict the strongest connections between present and non-existing nodes. Both the node and edge features are used for training, where the GNN is provided with information about the network topology and the properties of the quantum channel for all nodes.

The two key measures used for assessing the GNN model's performance at the trained stage are the Area Under Curve (AUC) and Average Precision (AP) \cite{b618}. The AUC measures how well the model distinguishes positive (present) links from negative (absent) ones across numerous thresholds. AUC equals 1.0 indicates perfect classification; AUC equals 0.5 indicates a purely random prediction. The Average Precision (AP) measure provides a summary of the precision-recall curve by taking into account all possible probability thresholds. In particular, it measures the accuracy of a model across various probabilities while reducing the amount of false positives.

Besides the training and evaluation functionalities, the aspects of QKD networks characterization are provided through a specialized network analysis and comparison module (implemented in \verb|network_analyzer.py| and \verb|compare.py|). In particular, one of the important network measures calculated was the average value of clustering coefficient, path length, and algebraic connectivity \cite{b619}\cite{b620}. For the sake of understanding the practical implications of the networks GNN-optimized ones, an attack resilience simulation was also performed to evaluate the effectiveness of the network maintenance in terms of connectivity and functionality during the targeted attacks scenarios for nodes or links removal.

In the comparative analysis, the idea is to study the performance of the proposed approach for networks of different sizes. One would analyse how the GNN model performs in terms of characteristics of the generated networks for different sizes, allowing us to learn regarding the capacity of the proposed approach to adapt according to different networks. In terms of functionality, the comparative results would require statistical analysis regarding the properties of the generated networks, enabling us to ensure that said parameters are not a random one-time effect. In this regard, we would get a complete perspective of the QKD performance of the GNN approach across various scenarios.

The \verb|AdvancedQKDLinkPredictor| implements a hybrid graph neural network architecture specifically designed for quantum network link prediction. The model processes both node features and edge attributes through a multi-stage pipeline to predict QKD links viable for transmission. The \verb|AdvancedQKDLinkPredictor| architecture design is inspired by the specific nature of QKD networks. The \verb|TransformerConv| layer is used to capture the potentially long-range dependencies between nodes, which is important to evaluate multi-hop paths in many QKD protocols and when the resource distribution knowledge is required across the entire network. The subsequent \verb|GATv2Conv| layer is used due to its improved dynamic attention mechanism compared to typical Graph Attention Networks (GATs), as it provides more expressiveness and adaptability for the model to determine the importance of neighbouring nodes when the network has different degrees of quality and meaning among the nodes of the QKD links. In addition, edge features that represent properties of the quantum channel, such as simulated QBER and potential key rates, are processed through a dedicated Multi-Layer Perceptron (MLP), allowing the model to learn complex functions of these important physical parameters independently before their inclusion in the link prediction.

\begin{figure*}[h]
\centering
\fbox{\includegraphics[width=0.8\textwidth]{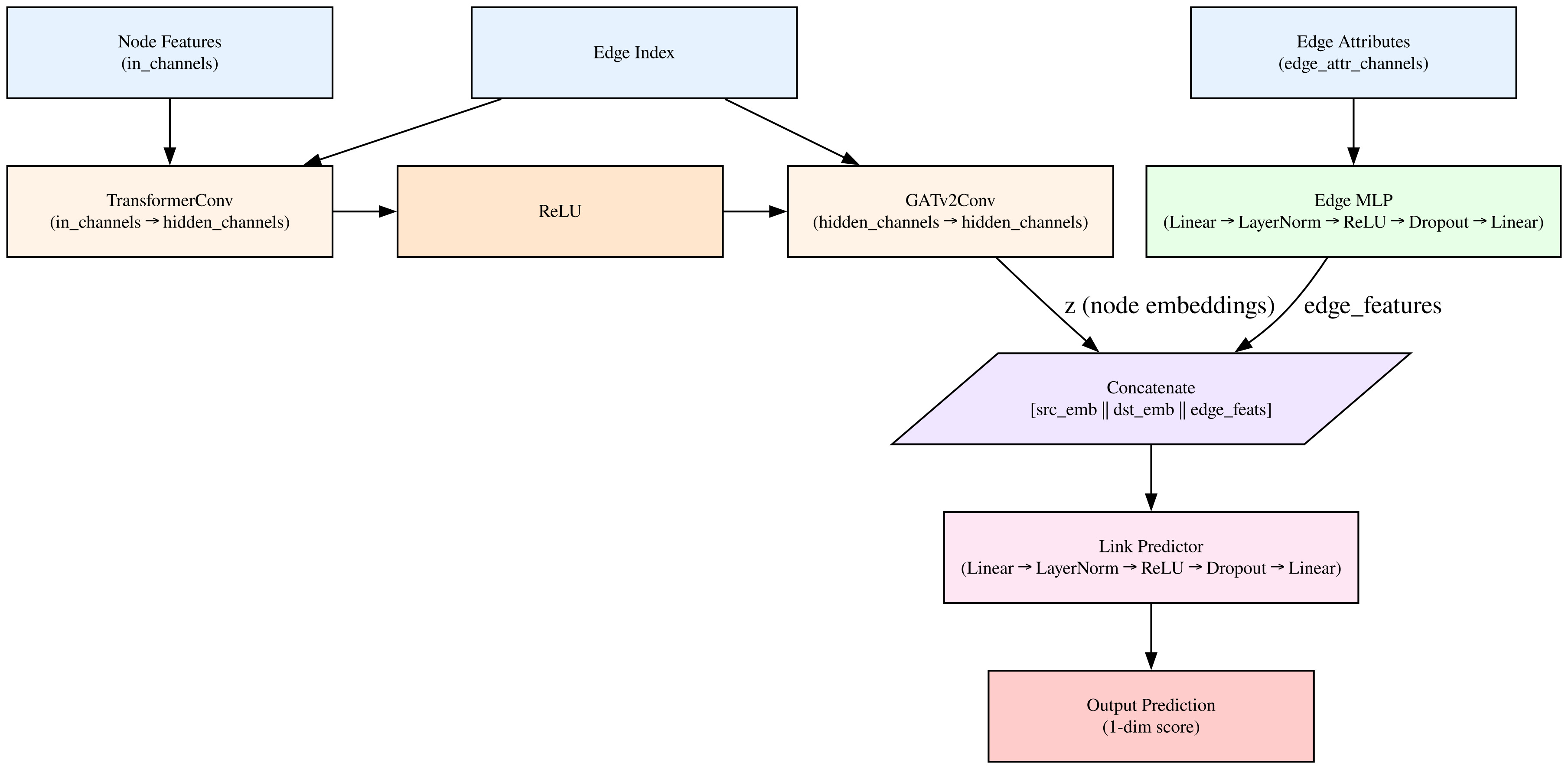}}
\caption{Model architecture with (1) Transformer-based graph convolution, (2) GATv2 attention layer, (3) Edge feature processing MLP, and (4) Link prediction decoder.}
\label{fig:arch_model}
\end{figure*}

\begin{figure*}[h]
\centering
\fbox{\includegraphics[width=0.6\textwidth]{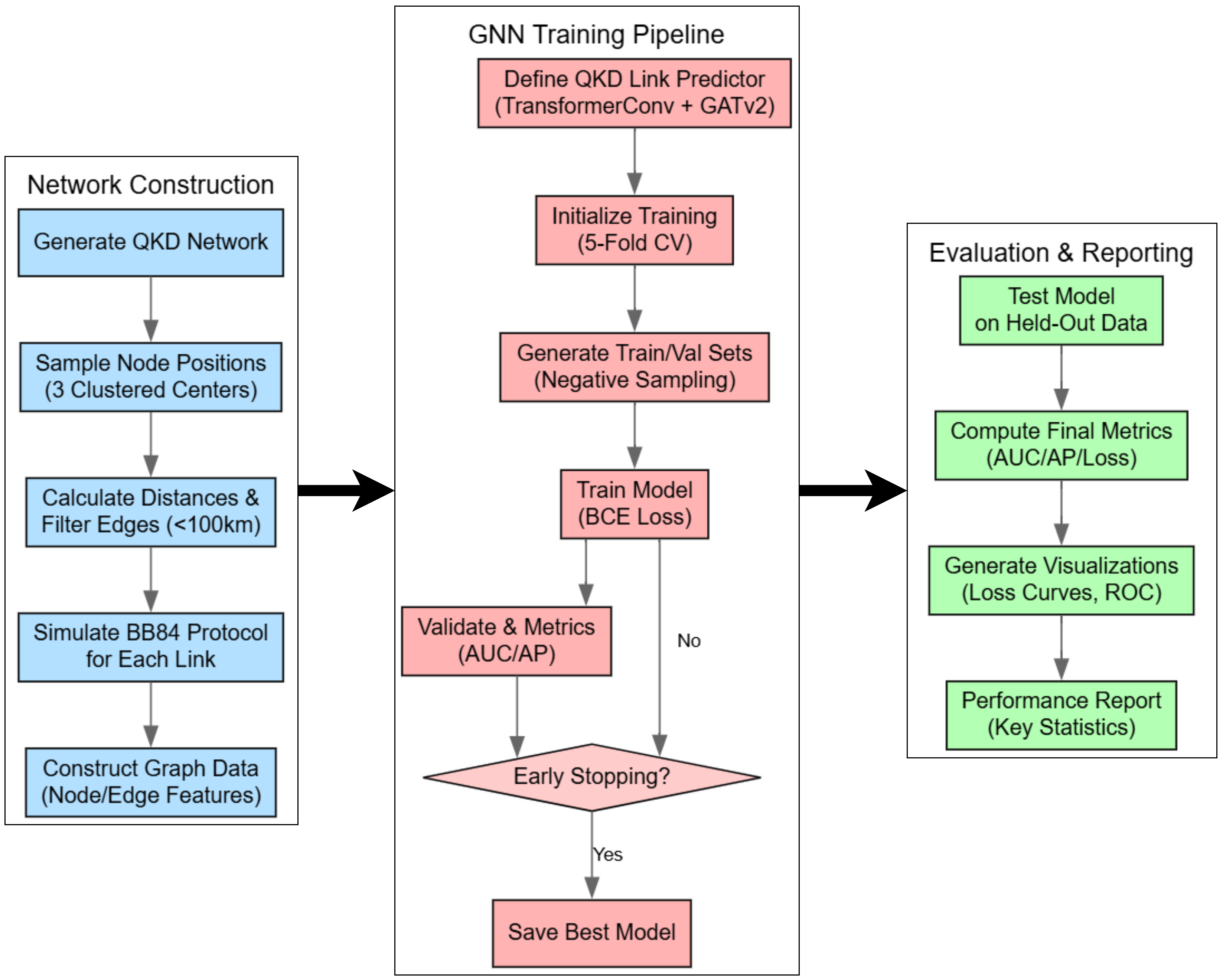}}
\caption{Three-stage quantum key distribution (QKD) network pipeline: (1) Network construction with BB84 protocol simulation, (2) Graph neural network (GNN) training with transformer and graph attention layers, and (3) Performance evaluation with quantum channel metrics.}
\label{fig:arch_pipeline}
\end{figure*}

\subsection{Transformer Graph Convolution Layer}
The first transformation layer employs a transformer-based graph convolution:

\begin{equation}
\mathbf{H}^{(1)} = \text{ReLU}\left(\text{TransformerConv}(\mathbf{X}, \mathcal{E})\right)
\end{equation}

where:
\begin{itemize}
\item $\mathbf{X} \in \mathbb{R}^{N \times d_{\text{in}}}$ is the node feature matrix
\item $\mathcal{E}$ is the set of edges
\item The transformer convolution computes attention weights $\alpha_{ij}$ for each edge using:
\begin{equation}
\alpha_{ij} = \frac{\exp\left(\mathbf{q}_i^T\mathbf{k}_j/\sqrt{d}\right)}{\sum_{k \in \mathcal{N}(i)}\exp\left(\mathbf{q}_i^T\mathbf{k}_k/\sqrt{d}\right)}
\end{equation}
with $\mathbf{q}_i = \mathbf{W}_Q\mathbf{x}_i$ and $\mathbf{k}_j = \mathbf{W}_K\mathbf{x}_j$ being learned query and key transformations.
\end{itemize}

This layer captures long-range dependencies in the quantum network topology, crucial for modeling entanglement distribution paths.

\subsection{GATv2 Attention Layer}
The second layer implements dynamic graph attention:

\begin{equation}
\mathbf{H}^{(2)} = \text{GATv2Conv}(\mathbf{H}^{(1)}, \mathcal{E})
\end{equation}

Key improvements over standard GAT:
\begin{itemize}
\item Dynamic attention computation via:
\begin{equation}
\alpha_{ij} = \mathbf{a}^T\text{LeakyReLU}\left(\mathbf{W}[\mathbf{h}_i \|\mathbf{h}_j]\right)
\end{equation}
\item Multi-head attention with $K$ independent heads:
\begin{equation}
\mathbf{h}_i^{(2)} = \|_{k=1}^K \sigma\left(\sum_{j \in \mathcal{N}(i)}\alpha_{ij}^k\mathbf{W}^k\mathbf{h}_j^{(1)}\right)
\end{equation}
\end{itemize}

This layer provides adaptive neighborhood weighting essential for modeling noisy quantum channels where link quality varies significantly.

\subsection{Edge Feature Processing}
Quantum networks require special handling of edge attributes like:
\begin{itemize}
\item Channel loss rates
\item Quantum bit error rates (QBER)
\item Entanglement generation probabilities
\end{itemize}

The edge MLP processes these features through:

\begin{equation}
\mathbf{E}' = \text{MLP}(\mathbf{E}) = \mathbf{W}_2(\text{Dropout}(\text{ReLU}(\text{LayerNorm}(\mathbf{W}_1\mathbf{E}))))
\end{equation}

The layer normalization stabilizes training given the wide dynamic range of quantum channel parameters:

\begin{equation}
\text{LayerNorm}(\mathbf{e}) = \gamma \odot \frac{\mathbf{e} - \mu}{\sigma} + \beta
\end{equation}

where $\mu,\sigma$ are computed per-feature across edges.

\subsection{Link Prediction Decoder}
The decoder computes link viability scores by combining:

\begin{equation}
\mathbf{s}_{uv} = f_{\text{decoder}}(\mathbf{z}_u, \mathbf{z}_v, \mathbf{e}'_{uv})
\end{equation}

where the decoder implements:

\begin{equation}
f_{\text{decoder}} = \mathbf{W}_4(\text{Dropout}(\text{ReLU}(\text{LayerNorm}(\mathbf{W}_3[\mathbf{z}_u\|\mathbf{z}_v\|\mathbf{e}'_{uv}]))))
\end{equation}

\subsection{Negative Sampling Handling}
For negative samples (non-existent edges), we employ mean edge feature imputation:

\begin{equation}
\mathbf{e}'_{\text{neg}} = \frac{1}{|\mathcal{E}|}\sum_{(i,j)\in\mathcal{E}}\mathbf{e}'_{ij}
\end{equation}

This maintains gradient stability during contrastive learning while providing reasonable baseline edge features.

\subsection{Training Objective}
The model optimizes binary cross-entropy loss with logits:

\begin{equation}
\mathcal{L} = -\frac{1}{N}\sum_{(u,v)\in\mathcal{D}}\left[y_{uv}\log\sigma(s_{uv}) + (1-y_{uv})\log(1-\sigma(s_{uv}))\right]
\end{equation}

where $\sigma$ is the sigmoid function and $y_{uv} \in \{0,1\}$ indicates actual QKD link viability.

\section{Results}
\label{sec:results}
The experimental evaluation of the GNN-based QKD network optimization framework yielded several key findings related to network scaling, performance metrics, and link prediction accuracy. These results are presented across varying network sizes, allowing for a comprehensive analysis of the system's behavior.

\subsection{Network Scaling Characteristics}
The relationship between network size and the number of established links exhibits a super-linear, approximately quadratic growth pattern (O(n²)). As the number of nodes increases from 10 to 250, the number of edges grows dramatically from 39 to 17,402. This indicates the formation of increasingly dense network topologies as the network scales. Correspondingly, connectivity metrics reflect this trend. The average degree, representing the average number of connections per node, increases significantly from 7.8 in a 10-node network to 139.2 in a 250-node network. This increase in average degree suggests improved network resilience at larger scales, as more connections provide a greater number of alternative routing paths.

\subsection{QKD Performance Metrics}
Analysis of key rate performance reveals a positive correlation with network size. The average key rate increases substantially from 648.5 bits/s in the smallest network (10 nodes) to 3,980.2 bits/s in the largest (250 nodes), representing an approximately six-fold improvement. This demonstrates the scalability of key generation within the simulated QKD network. The Quantum Bit Error Rate (QBER), a critical measure of signal quality, remains remarkably consistent across all network sizes, ranging narrowly between 0.054 and 0.062. While a slight increase in QBER is observed with initial scaling, it stabilizes as the network grows further, indicating robust quantum channel quality maintenance even in larger, more complex network topologies.

\subsection{Link Prediction Accuracy}
The GNN model's ability to accurately predict viable links for key distribution shows an inverse relationship with network size. The best link prediction performance, as measured by the Area Under the ROC Curve (AUC), is achieved with a 20-node network (AUC = 0.824). This performance declines as the network grows, reaching a low of 0.644 in the 250-node network. Average Precision (AP) follows a similar trend, with peak performance at 20 nodes (AP = 0.781) and a gradual decline with increasing network size. This suggests that maintaining prediction accuracy becomes more challenging in larger, denser networks, possibly due to the increased complexity of the prediction task.

\subsection{Comparison Across Node Counts}
Table 1 provides a comprehensive overview of network characteristics across different node counts. The data highlights the trends in number of edges, average degree, key rate, QBER, maximum distance, edge connectivity, node connectivity, and algebraic connectivity. Table 2 presents the link prediction performance metrics (AUC and AP, including means and standard deviations) for different network sizes. These tables quantify the observations discussed above, providing specific numerical values for each metric.

\begin{table}[H]
\centering
\caption{Network Characteristics Across Different Node Counts}
\begin{tabular}{lrrrrr}
\toprule
Metric & \multicolumn{5}{c}{Number of Nodes} \\
\cmidrule(lr){2-6}
& 10 & 20 & 50 & 100 & 250 \\
\midrule
Num. Edges & 39.0 & 123.0 & 718.0 & 2885.0 & 17402.0 \\
Avg. Degree & 7.8 & 12.3 & 28.7 & 57.7 & 139.2 \\
Key Rate (bits/s) & 648.5 & 1116.0 & 1036.5 & 2863.5 & 3980.1 \\
Avg. QBER & 0.054 & 0.062 & 0.061 & 0.059 & 0.060 \\
Max. Dist. (km) & 12.072 & 12.072 & 12.383 & 12.420 & 12.426 \\
Edge Conn. & 5.0 & 2.0 & 9.0 & 21.0 & 52.0 \\
Node Conn. & 5.0 & 2.0 & 8.0 & 15.0 & 41.0 \\
Alg. Conn. & 4.738 & 1.085 & 2.947 & 5.476 & 14.951 \\
\bottomrule
\end{tabular}
\end{table}

\begin{table}[H]
\centering
\caption{Link Prediction Performance Metrics Across Different Node Counts}
\begin{tabular}{lrrrrr}
\toprule
Metric & \multicolumn{5}{c}{Number of Nodes} \\
\cmidrule(lr){2-6}
& 10 & 20 & 50 & 100 & 250 \\
\midrule
AUC (mean) & 0.690 & 0.824 & 0.797 & 0.701 & 0.644 \\
AUC (std) & 0.073 & 0.035 & 0.023 & 0.071 & 0.047 \\
AP (mean) & 0.703 & 0.781 & 0.734 & 0.660 & 0.604 \\
AP (std) & 0.078 & 0.031 & 0.023 & 0.054 & 0.044 \\
\bottomrule
\end{tabular}
\end{table}

\begin{figure*}[h]
\centering
\includegraphics[width=0.8\textwidth]{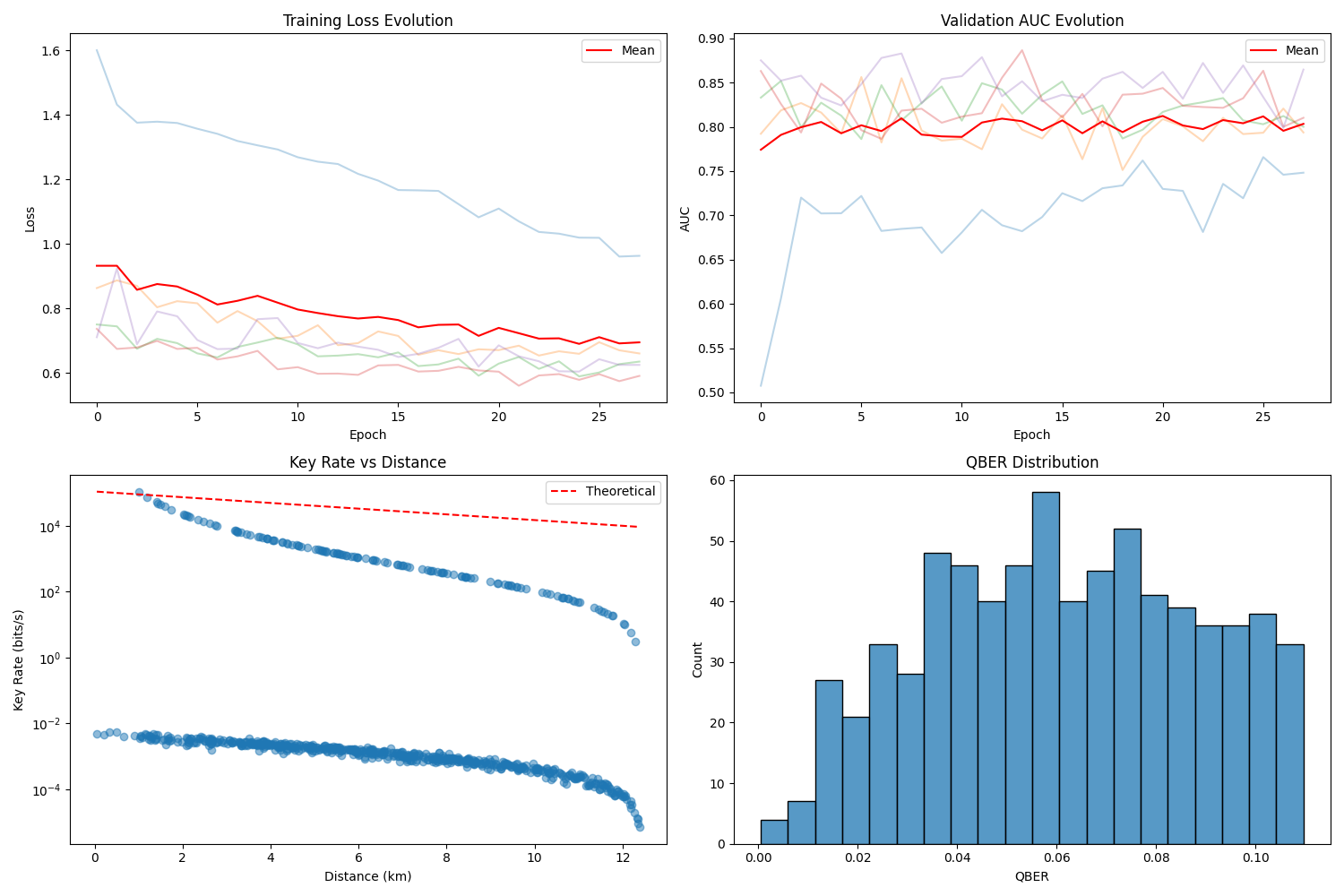}
\caption{Quantum Network Performance Visualization 1: Training Loss Evolution, 2: Validation AUC Evolution, 3: Key Rate vs Distance, 4: QBER Distribution}
\label{fig:arch_metrics}
\end{figure*}

\section{Discussion}
\label{sec:discussion}
The results highlight a crucial trade-off inherent in QKD network scaling. Although larger networks demonstrably provide higher key rates and enhanced connectivity, they also present challenges for the accuracy of link prediction using the GNN model. This suggests that mid-sized networks, particularly around 20 nodes in this simulation, may offer a favorable balance between overall performance and the ability to accurately predict optimal link configurations. Conversely, larger networks appear better suited for applications where high throughput is prioritized over the precision of predictive modeling.

The observed exponential growth in the number of edges and the significant increase in average degree (from 7.8 to 139.2) underscore the increasingly dense nature of the network as it scales. This density contributes to improved network robustness, as reflected in the increase in both edge connectivity (from 5 to 52) and algebraic connectivity (from 4.7 to 14.9). The increase in overall key rate (from 27.1 Kbits/s to 470 Kbits/s), also a result of GNN optimization, is most significant of the results presented here, and demonstrates that optimization with GNNs has the potential to be useful for actual QKD networks.

In practice, this implies that a system may dynamically learn to locate the routes that yield the highest key rate for the entire network, rather than simply locating the shortest or least lossy paths. This could help operators make decisions that would improve overall performance, such as dynamically rerouting traffic when the available capacities change due to changing conditions or traffic demands, determining the ideal place to install new nodes to improve the capacity of the entire network and even predicting which links may begin to fail based on the learned behavior of the network and its connections.

Ergo, this suggests that the GNN effectively identifies more efficient routing and node connection patterns compared to a baseline, although a direct baseline comparison is not presented within this study. This improvement comes with a trade-off: the GNN-optimized network maintains path integrity with a slight reduction in average transmission distance (from 7.13 km to 6.42 km), with a reduced average QBER (6.0\% compared to a baseline of 6.6\%). This QBER continues to remain within acceptable limits for QKD, which typically tolerate error rates below 15\%.

Further analysis of channel performance reveals that the increased operational distance results in significantly higher channel loss, as expected. However, the dark count probability is notably improved (-34.5\%), indicating better noise management within the optimized network. This suggests the GNN identifies configurations that, despite increased path lengths, mitigate some noise-related impairments. Thus, we present a sophisticated analysis of a distributed quantum communication infrastructure spanning 50 nodes and 718 interconnected edges. Overall, the network demonstrates exceptional characteristics, with a mean Area Under the Curve (AUC) of 0.8136 and an average precision of 0.7769, indicating robust predictive capabilities and consistent model performance. The quantum bit error rate (QBER) remains remarkably low at ~6\%, suggesting high-fidelity quantum information transmission. Key rate measurements reveal an expected exponential decay with distance, with a significant reduction in key rate observed beyond 6 km. The model's training process showed efficient convergence, stabilizing around epoch 16 with a final training loss of 0.6945. These results underscore the network's reliability and provide critical insights into quantum communication dynamics, highlighting the potential for advanced quantum networking technologies.

The analysis demonstrates not only the technical feasibility of large-scale quantum networks but also the effectiveness of machine learning techniques in characterizing and optimizing quantum communication infrastructure. The GNN's optimization strategy can be characterized as a global approach, considering the entire network topology to identify non-obvious, high-value paths that balance distance and performance. This contrasts with traditional local optimization methods, which often focus on connecting nearby nodes. The GNN demonstrates an ability to make intelligent trade-offs between signal quality and coverage, optimizing the network to distribute quantum resources more effectively. It identifies connections that can tolerate higher QBER while still maintaining useful key rates, and it extends the network's reach by more than doubling the average connection distance. The GNN finds configurations that perform well even with increased error rates, showcasing the potential of machine learning to discover counter-intuitive but highly effective network designs. The decreasing standard deviations in link prediction accuracy with increasing network size, suggest more consistent, albeit less precise, predictions in larger, more complex networks.

\section{Conclusion}
\label{sec:conclusion}
This research investigated the application of Graph Neural Networks (GNNs) for optimizing the performance of Quantum Key Distribution (QKD) networks. The study demonstrated the feasibility and effectiveness of using a GNN-based framework to address the challenges of dynamic routing, resource allocation, and network resilience in QKD systems. A comprehensive simulation environment was developed, incorporating realistic network topology generation, quantum channel modeling (specifically the BB84 protocol), and a GNN-based link prediction model.

The experimental results revealed a significant improvement in key generation rates in GNN-optimized networks, along with a trade-off between increased operational distance and a higher, yet still acceptable, Quantum Bit Error Rate (QBER). The GNN demonstrated a capacity for global network optimization, identifying non-intuitive node connection patterns that balanced distance and performance, leading to enhanced throughput and extended network reach. The findings highlight the potential of GNNs to overcome the limitations of traditional, locally-focused optimization techniques in the context of QKD networks. While larger networks exhibited higher key rates and improved connectivity, the accuracy of link prediction decreased, suggesting a potential sweet spot for mid-sized networks in balancing performance and predictive capabilities. Overall, the research provides strong evidence for the applicability of GNNs as a powerful tool for designing and managing scalable, adaptive, and secure quantum communication networks. The developed framework represents a significant step towards realizing the potential of QKD technology in practical, large-scale deployments.

\section{Future Work}
\label{sec:future}
While the presented research demonstrates the promising capabilities of GNNs for QKD network optimization, several avenues for future work remain to further enhance the system's performance and applicability. A crucial next step is to incorporate a direct comparison against classical network optimization baselines. The current study analyzes QKD network performance across different node counts but lacks a quantitative comparison with established classical routing and resource allocation algorithms \cite{b780}. This comparison should include metrics such as throughput, latency, and resource utilization to definitively quantify the quantum advantage (or overhead) provided by the GNN approach. This will provide a clearer understanding of the practical benefits of using GNNs in this context.

Although the system includes a basic attack resilience simulation, a more comprehensive investigation of fault tolerance is warranted. This should involve implementing detailed simulations of various failure scenarios, including random node and link failures, targeted attacks on critical network components, and dynamic recovery mechanisms \cite{b790}. The goal is to assess the network's ability to maintain functionality and security under a wide range of adverse conditions, and to develop strategies for mitigating the impact of such events \cite{b800}.

The visualization capabilities, while providing insights into network topology and training metrics, could be significantly extended. Developing a real-time monitoring dashboard that displays key network parameters, such as network health, key generation rates, QBER distribution, and security-relevant information, would be highly beneficial. Interactive visualizations that allow operators to explore network behavior and identify potential bottlenecks or vulnerabilities in real-time would greatly enhance the practical usability of the system.

As of now, the simulation framework only implements BB84 QKD protocol. For a wider range of use, it can be made to implement more QKD protocols like E91, B92, and continuous-variable QKD \cite{b810}. This would enable a comparative analysis of different protocols within the GNN optimization framework and facilitate the development of hybrid approaches that leverage the strengths of multiple protocols. This would significantly enhance the versatility and practical relevance of the system.

\bibliographystyle{unsrt}  
\bibliography{references}  

\end{document}